\documentclass[10pt,journal,compsoc]{IEEEtran}
\ifCLASSOPTIONcompsoc
  \usepackage[nocompress]{cite}
\else
  \usepackage{cite}
\fi

\usepackage{xspace}
\usepackage{booktabs, multirow}
\usepackage{setspace}
\usepackage{amsmath,amssymb,array,threeparttable,multirow,amsthm}
\usepackage{graphicx, subfigure}
\usepackage{float}

\usepackage{graphicx}
\usepackage{epsfig,endnotes}
\usepackage{algorithm}
\usepackage[noend]{algpseudocode}
\usepackage[lettersize]{lscape}
\usepackage{breakurl}

\usepackage[shortlabels]{enumitem}
    \setlist[enumerate, 1]{1.}
    \setlist{leftmargin=5.0mm}
\usepackage{tikz}
\usepackage{epstopdf}
\usepackage{adjustbox}
\usepackage{romannum}

\usepackage{color}
\usepackage{soul}

\newlist{Steps}{enumerate}{1}
\setlist[Steps,1]{label=Step~\arabic*.,leftmargin=*}

\begin{document}
\title{Barriers in Seamless QoS for Mobile Applications}

\author{Mohammad A. Hoque, Hassan Abbas, 
        Tong Li, 
        Yong Li, 
        Pan Hui, 
        and 
        Sasu Tarkoma
        \IEEEcompsocitemizethanks{
        \IEEEcompsocthanksitem 
        Mohammad A. Hoque, Hassan Abbas, Pan Hui, and Sasu Tarkoma are with the University of Helsinki, Finland (e-mail: firstname.lastname@cs.helsinki.fi).
        \IEEEcompsocthanksitem 
        Tong Li and Pan Hui are with the Honkg Kong University of Science and Technology, Gong Kong (e-mail: t.li@connect.ust.hk,  
panhui@cse.ust.hk).
        \IEEEcompsocthanksitem 
        Yong Li is with the Tsinghua University, China (e-mail: liyong07@tsinghua.edu.cn).

		\IEEEcompsocthanksitem 
		Mohammad A. Hoque is the corresponding author of this paper.
}
}



\IEEEtitleabstractindextext{%
\begin{abstract}
For seamless QoS, it is important that all the stakeholders, such as the hosts, applications, access networks, routers, and other middleboxes, follow a single protocol and they trust each other. In this article, we investigate the participation of these entities in providing QoS over wireless networks in light of \emph{DiffServ} QoS architecture. We initiate the study by investigating WiFi and Cellular network traces, which further motivates us a thorough investigation of these stakeholders with empirical measurements.  Our findings are the followings. (i) Modern mobile VoIP applications request QoS to the network. (ii) While the operating systems support basic APIs requesting QoS, application developers either are not aware of such requirements or they misuse the architecture for improved QoS. (iii) Wireless access networks rewrite the QoS requirements at the edge and enforce the rest of the routers or hops to provide best effort service. (iv) QoS requests are also nullified by the secure tunnels. (v) Although the access networks may nullify the QoS requests, the performance of the  network still may differentiate traffic. We further emphasize that although the latest 5G network considers \emph{DiffServ} QoS framework, it cannot deal with the challenges posed by the privacy related applications. Network neutrality is going to pose similar challenges.
\end{abstract}

}

\maketitle

\IEEEdisplaynontitleabstractindextext

\IEEEpeerreviewmaketitle



\section{Introduction}

Mobile devices have been generating 60\% of all Internet traffic and this share will increase by five-fold in 2021. Cisco Visual Network index suggests that traffic from various services such video broadcast,  live streaming, AR/VR applications will grow by a five-to-seven fold from 2016 to 2021~\cite{cisco2017}. Mobile applications have been evolving very fast. Different applications may have different QoS requirements from networks. For a file download from the Internet, it is good to have high bandwidth. Since the file cannot be used until the download is complete, the latency is not crucial.  On the contrary, low latency is crucial for VoIP applications so that end users do not experience a distorted voice, and thus in this case latency is the higher priority.

\emph{DiffServ} (RFC~2474~\cite{Nichols:1998}) architecture has been the long standing match for the Internet to deploy different QoS \emph{policies} or \emph{behavior} with simple mechanisms in a simple to complex network compositions. The reason is that it operates at the IP layer, and therefore, all the network entities working at the IP layer can easily recognize and set the QoS policies. Such entities include mobile devices, host machines with network cards, proxy servers, routers and switches. At the end hosts, QoS policies can be set by the end users or applications. Nevertheless, it is not well studied, whether the mobile applications are taking advantage of this architecture, or they are facing other challenges.


In this article, we first investigate the presence of QoS requests in a WiFi and a commercial cellular network traces. The datasets contain more than 750M flows from various applications and only 2\% of the flows were marked with Differentiated Service Code Point (DSCP) values. This finding has been surprising for us and motivates us to  perform a systematic study with Android and iOS applications communicating over Wi-Fi and Cellular networks.

The study reveals the contribution of all network elements in providing QoS for a end-to-end communication in the WiFi or cellular networks while using a number of mobile applications requiring QoS from the networks.We further depend a simple \emph{DiffServ}-based QoS measurement tool.  The tool can simulate various traffic patterns. It sets DSCPs at the application layer in mobile devices for different traffic patterns and network settings, and analyzes the corresponding traffic at the server. It is important for the end users that they understand the performance of their applications, operators' policies, and the implication of using network control or management software, such as VPN clients, for different applications. Our findings are the followings. 

\begin{itemize}
\item A few mobile applications mark flows with QoS requirements irrespective of the network type. In other words, mobile operating systems support APIs to allow the applications to tag the flows with DSCPs. 
\item However, the developers either do not take the advantage of  APIs or misuse for improved QoS. 
\item Nevertheless, when the packets enter into the wireless networks, these edge networks may rewrite or nullify the requests.  
\item The QoS requests are also nullified when the end users use some secure tunnels or Virtual Private Network (VPN) applications.  
\item Our experiments for the intermediary routers suggest that they forward the requests as they are and adhere the QoS policies requested by the applications.
\end{itemize}

This study at high level reveals that the decentralized nature of \emph{DiffServ} hardens the synchronization of the network elements in providing true QoS, as different players can act independently. It also shows that the service providers have been giving their best efforts to serve various applications at present. Although such effort has been sufficient for the present applications, with the advent new technologies a new set of latency sensitive or bandwidth consuming applications, such as augmented reality, virtual reality, medical treatment, and  remote surgery, are emerging. Therefore, 3GPP working group is addressing these challenges with new radio technologies, spectrum or bandwidth aggregation in the wireless networks, and network slicing in the core networks~\cite{7926920} or even slicing the radio networks~\cite{8253541}. Nevertheless, it is important that traffic from these applications should be treated according to the required and similar manner by all the intermediary stakeholders to meet the application or end user demand or QoS. At the same time 3GPP is also considering \emph{DiffServ} for QoS management in 5G networks. Although these new technological advancements are laying out the foundation to meet the QoS goals, we believe that political issues, such as net neutrality, will also play major roles in providing the guaranteed QoS.

%

The rest of the paper is organized as follows. Section 2 provides an overview of \emph{DiffServ} code points and their implications for QoS. In Section 3, we present our  findings 
from existing WiFi and cellular network traces. Section 4 discusses the performance of  various networks with various QoS requests and implications for 5G.  In Section 5, we discuss the implications of net neutrality for the QoS of present mobile applications from related work. Section 6 discusses the challenges for the upcoming 5G networks before concluding the paper. 

\section{DiffServ and DSCP}

\begin{figure}[h]
\centering
\includegraphics[width=0.95\columnwidth]{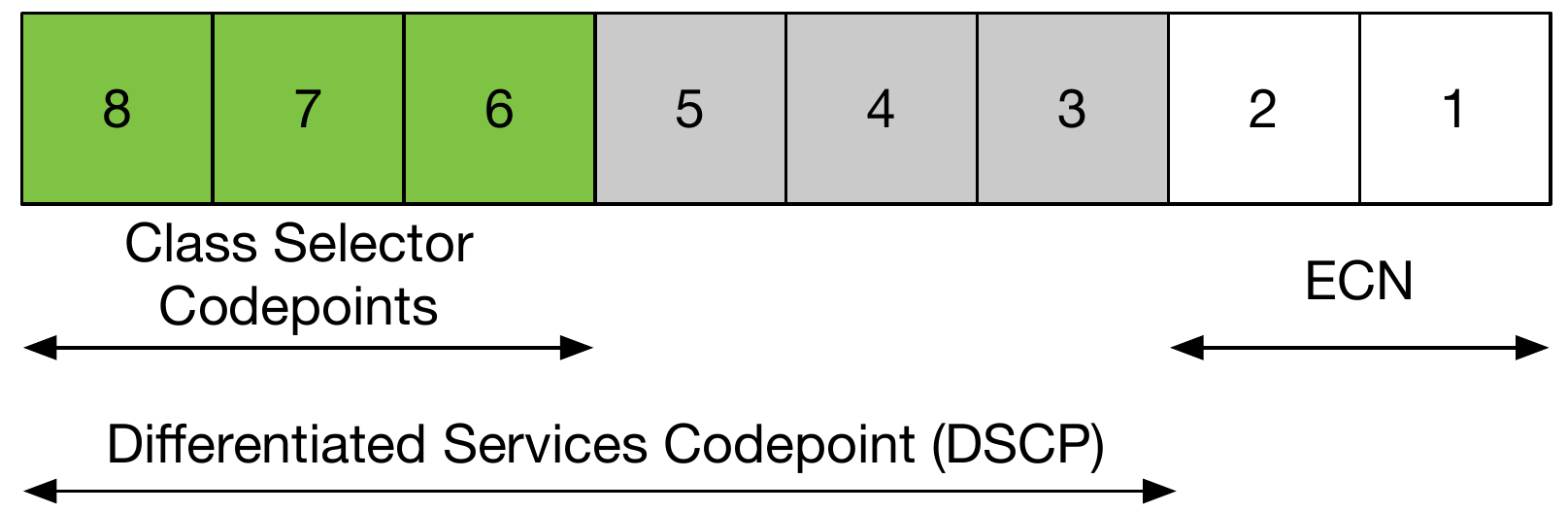}
\caption{Type of Service (ToS) or Traffic Class byte of the IPv4 or IPv6 Header. }
\label{fig:trafficlass}
\end{figure}

The \emph{DiffServ} architecture employs coarse-grained packet-based QoS. It consists of domains and each domain implements edge and core functionalities. The edge router implements complex traffic classification and marks the packets accordingly using DSCPs. These codes also define the activity of core routers and the treatment to be applied through scheduling and queue management policies. Figure~\ref{fig:trafficlass} illustrates that DSCP values occupy the most significant 6 six bit of the Type of Service (ToS) field of IPv4/IPv6 header. The most significant 3 bits act as the class selector, there are seven class selectors~\cite{Nichols:1998}, and maintain backward compatibility.  


\vspace{1mm}
\noindent\textbf{Default Behavior (CS0).} The default behavior is defined as CS0 and all the bits of DSCP are set 0. This class reflects the default behavior or emphasizes the best effort service where the congestion and loss are controlled on the Internet. 

\vspace{1mm}
\noindent\textbf{CS1-CS4.} With additional three bits, these four classes define Assured Forwarding (AFxy) policies for the nodes defined in RFC ~\cite{Geib:rfc2597}. The `x' subscript denotes the class. Each class is associated with a certain amount of buffer space and link bandwidth.  The `y' subscript denotes three drop precedences of the packets for a class. For example, if there is congestion on a link of a \emph{DiffServ} node, the packets with AF behavior classes will be dropped such that their drop probabilities are conditioned as $AFx1\leq AFx2\leq AFx3$. In other words, the packets with AFx3 will be dropped first. The participating nodes use them for various degree of forwarding guarantees. 

\vspace{1mm}
\noindent\textbf{CS5.} Some applications may need premium treatment from the network and Expedited Forwarding (EF) is defined in this class in RFC~\cite{Jacobson:1999:EFP:RFC2598}. Therefore, this class is suitable for real time applications such as VoIP, video conferencing, and trading applications which packets' require robust end-to-end treatment. Although EF policy can be configured with priority queuing and rate limiting, special attention should be paid, as all the packets cannot be treated according to the requirement in the case of congestion.

\vspace{1mm}
\noindent\textbf{CS6-CS7.} These two classes are reserved for exchanging networking protocol messages.

\begin{figure*}[t]
  \begin{center}
  \subfigure[\textsf{WiFi Traces}]{\label{fig:sigcommdscp}\includegraphics[width=.43\linewidth]{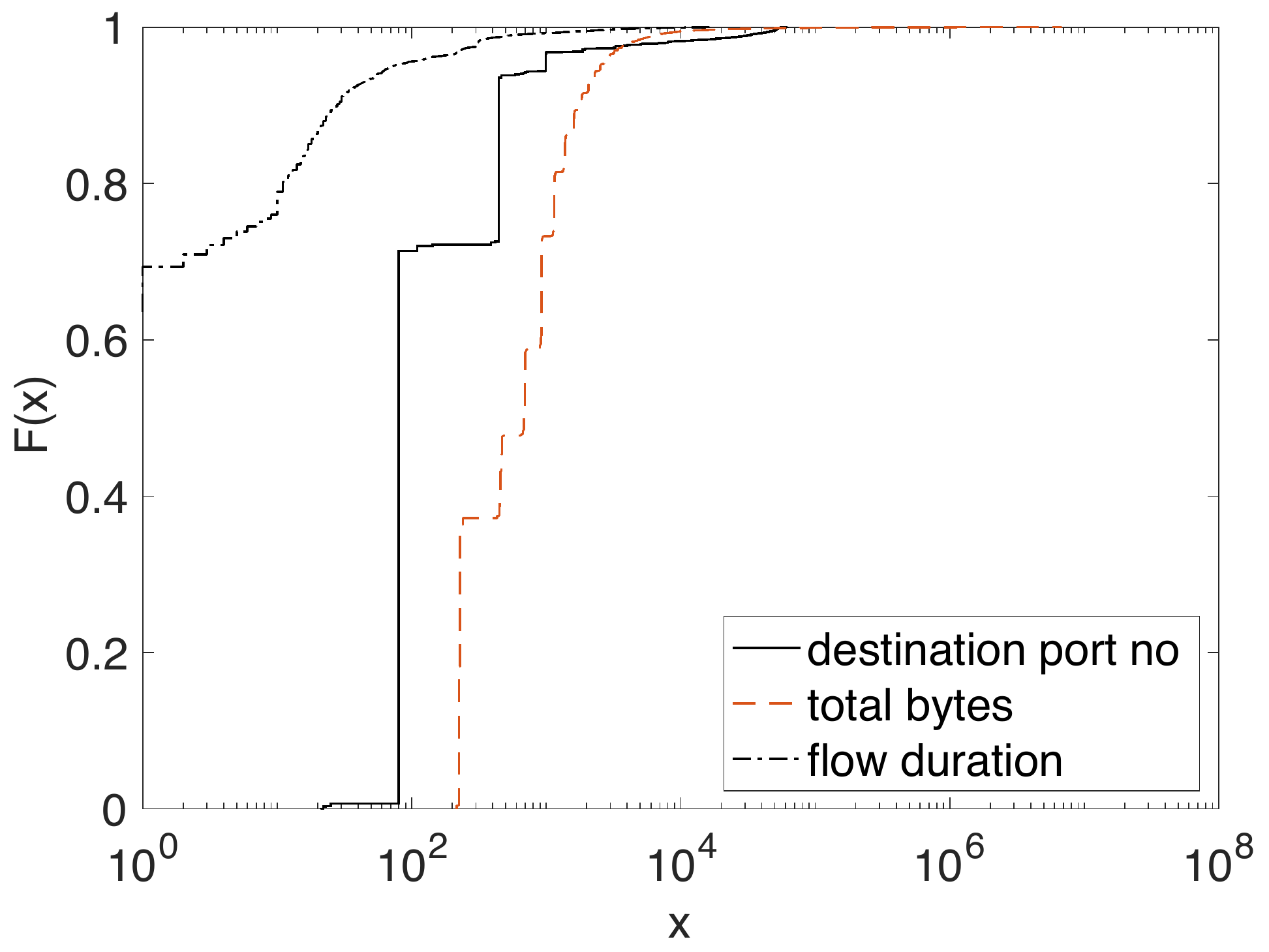}}
\subfigure[\textsf{Cellular Trace}]{\label{fig:DisC}\includegraphics[width=0.43\linewidth]{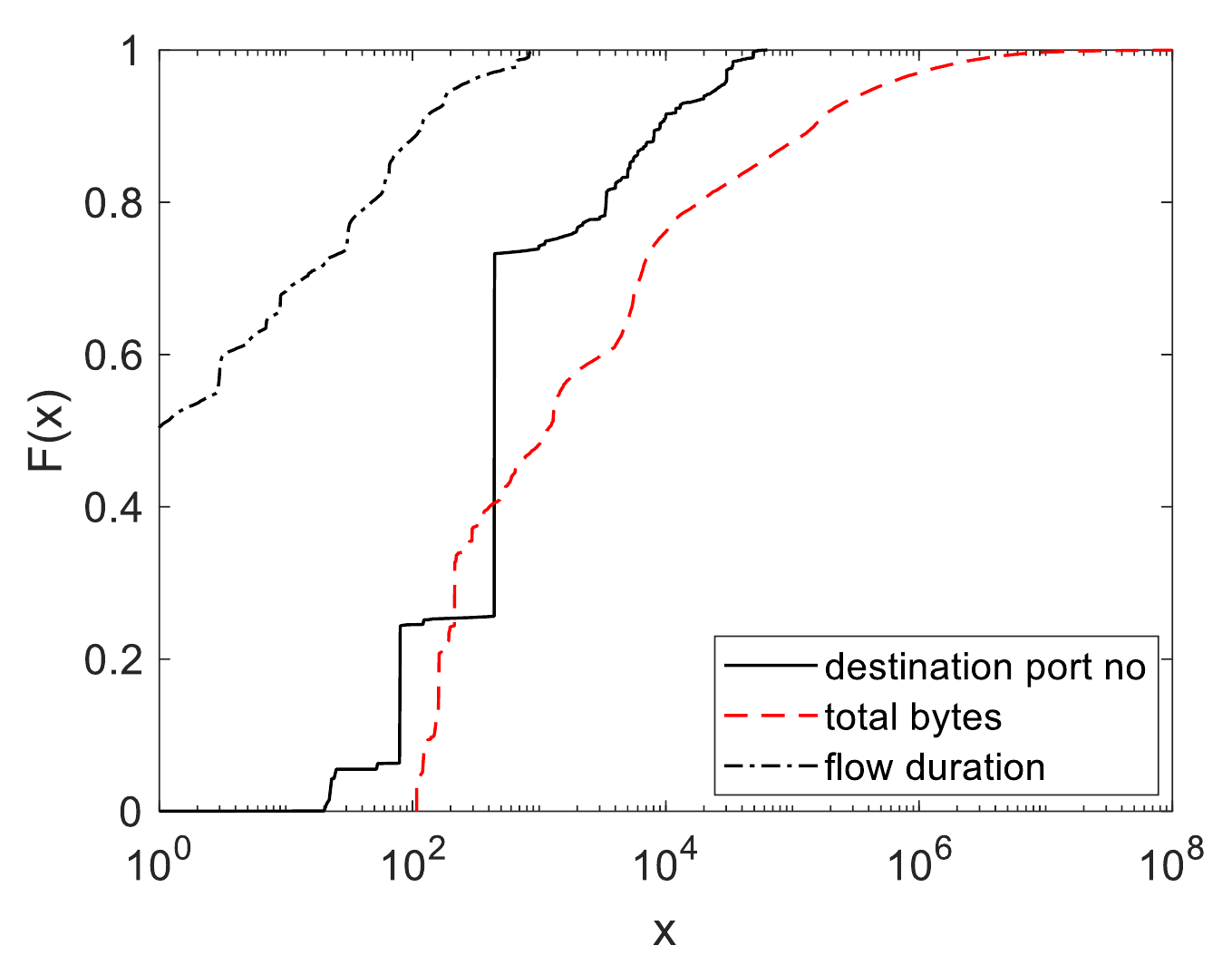}}
    \caption{The distribution of flow properties; port number, flow duration (seconds), and size (Bytes). The figures illustrate the properties of the flows marked with DSCPs other than CS0. The cdfs of the cellular network flows include the HTTP and HTTPS flows marked with ECN bits.}
    \label{fig:traces}
 \end{center}
 \end{figure*}

\vspace{1mm}
\noindent\textbf{ECN Bits.} While the DSCP signals the intermediary networks how to deal with the packets,  ECN, on the other hand, indicates the end points if there is congestion in the network and thus also indicates the reason for packet loss.  The ECN bits with either 01 or 10 combination expresses that the transport protocols at the end points are ECN capable. The 11 signals the end points about network congestion and the packets arriving to a router with a full queue will be dropped. 



\section{Wireless Network Traces}
In this section, we  analyze two traffic traces from WiFi and cellular networks. The first one is the popular Sigcomm 2008 WiFi trace~\cite{umd-sigcomm2008-20090302} collected during the conference in Seattle, USA. Next, we look into a commercial cellular network trace collected in 2014 by China Mobile, however, the dataset is not public. We analyze the flows  of various applications and protocol messages. A flow is a collection of packets between two end nodes defined by TCP/IP five-tuples.

\subsection{Sigcomm 2008 WiFi Traces}
This is one of the oldest WiFi datasets. Among 52 traces, we considered 24 WiFi traces. These traces had approximately 1 million flows. Even though more than two third of the flows were UDP, most of the TCP flows were carrying DSCP marks.  Only 10\% of the flows had DSCP or class selectors marked other than the default CS0.  These flows were marked with the basic class selectors, CS4, CS5, and CS6.  The markings  varied according to the applications and protocols. The mDNS and DNS flows had CS6 marked, whereas the ICMP messages were marked with CS5. On the other hand, SSH (port 22), mail client (port 993), HTTP (port 80), and HTTPS (port 443) flows were marked with CS4. Figure~\ref{fig:sigcommdscp} shows that HTTP traffic dominates. In the case of SSH traffic, we observed that DSCP varied based on flow directions. The SSH server marked flows with CS5 and the client marked  CS4. The figure also demonstrates that all the marked flows had only few bytes exchanged and their durations were quite small. We could not find any flow which would resemble VoIP or multimedia streaming traffic. This makes sense as these  applications were not popular at that time and such modern mobile applications were yet to be developed.

\subsection{Cellular Network Trace}

We also investigate the DSCP characteristics in a cellular network dataset. The dataset was collected from sample points of gateways in the core network by an Internet service provider. The dataset is over one week, namely 7 days. Regarding the privacy of subscribers, packet payloads are omitted, and IP addresses are anonymized. In addition, we look into the distribution of DSCP values except for CS0, as shown in Table \ref{table:DSCPdis}. We have found only 4\% of the flows were marked with DSCPs among 750M flows. Compared with other DSCP classes, CS6 takes the largest part and these flows are mostly DNS queries and ICMP messages.  We also observed some ICMP messages with CS7.  Some flows with unknown ports were marked with AF12 and AF21. Unlike the older WiFi traces, the HTTP and HTTPS flows were marked only with the ECN bits, rather than with any DSCP values.  Another observation is that some flows were marked with CS7. Similar to WiFi network flows, we can also see in Figure~\ref{fig:DisC} that such flows are of short durations and exchange few bytes, however, HTTPS dominates.


\begin{table}[h!]
\begin{center}
\caption{ Distribution of DSCP values except for 0 from an operator dataset collected over a period of seven days.}
{\footnotesize
 \begin{tabular}{| c |c|c|c|c|} 
 \hline
 \bf{DSCP value} &    AF11& AF21 & CS6 &CS7\\
 \hline
  Proportion &0.15\% & 0.41\% & 2.51\%&0.11\%\\\hline
 \end{tabular}} 
\label{table:DSCPdis}
\end{center}
\end{table}


\section{System, Application \& Network Support}
The Sigcomm WiFi traces reveal that a few applications had been using the class selectors at the dawn of mobile computing era. The cellular network traces exhibit similar behavior, except the usage of basic AF classes.

In order to understand the behavior of the entities involved in packet forwarding in end-end communication, we investigate already captured traffic, in 2018, of a number of modern mobile applications. This allows us to investigate the concern of the application developer towards QoS requirements. We further developed a set of simple socket based client-server applications with programming languages like Java and C, and performed a set of discrete tests. These applications send traffic from an end device to a remote server in Amazon cloud at some periodic interval,  e.g., 100ms and 1s. These tests reveal the behavior of APIs, Operating Systems, Application Emulators, Access Networks, Middleboxes, and Intermediary Routers. 

\subsection{System and Application Development} \label{system-performance}

We begin investigation with a desktop computer running Ubuntu operating system. In order to understand the application performance with DSCP, the application was initially developed in Java and we faced a daunting problem to set the DSCP value on a computer device. Java does not allow to set the traffic class because of a bug in the JVM, until some settings are tweaked. We resolved the problem by setting  \verb|java -Djava.net.preferIPv4Stack=true| JVM property. It concedes JVM to use the only IPV4 stack and not the IPV6 stack when interacting with the network layer. On the contrary, C provides better control over the socket options. 

\begin{figure*}[!ht]
\centering
\includegraphics[width=0.8\linewidth]{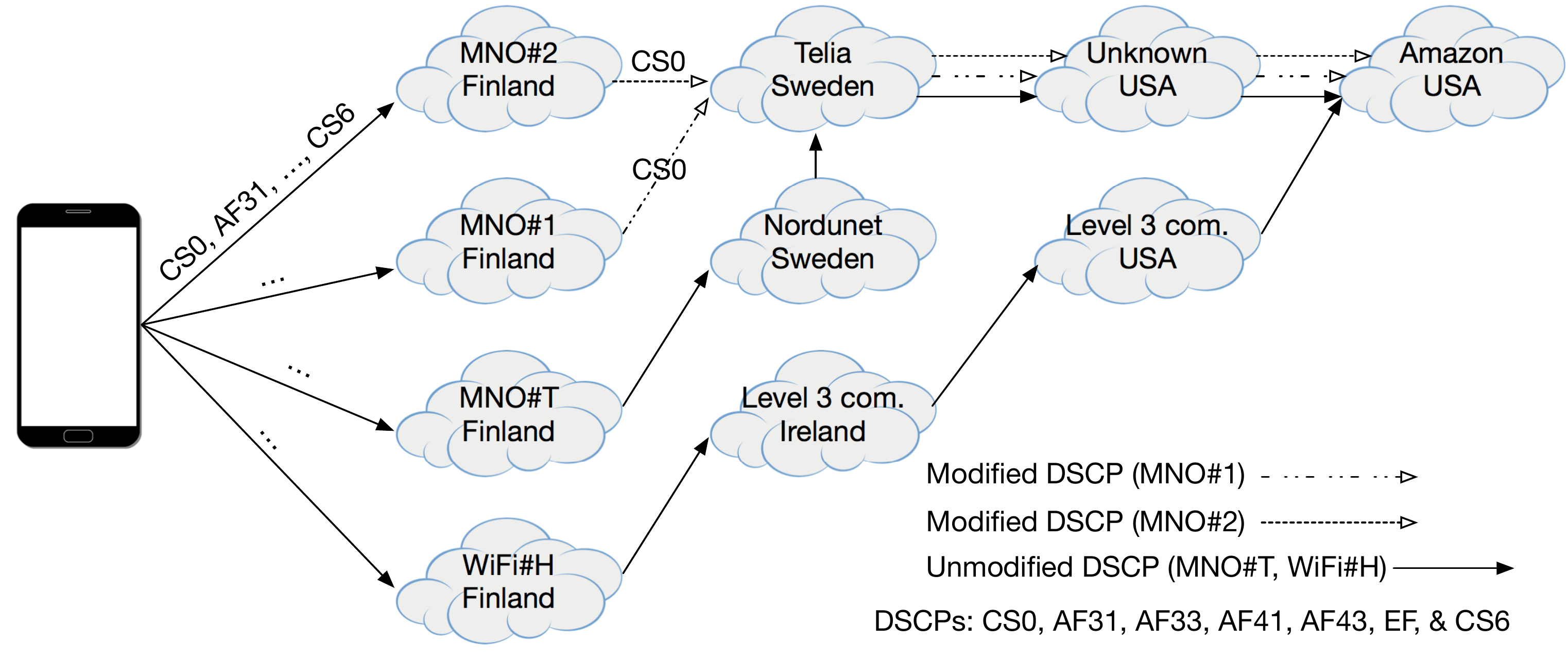}
\caption{The behavior of different networks in reseting the DSCP values. MNO\#1 and MNO\#2 are two public mobile networks. MNO\#T stands for the LTE test network in our research lab and WiFi\#H stands fore home WiFi gateway.}
\label{fig:dscpaths}
\end{figure*}

We use the socket API  \verb|socket.setTrafficClass()| for setting the DSCP traffic class of a flow from a mobile application. Similar to the desktop scenario, DSCP class setting did not work for Android Emulator either. However, we were able to set the desired traffic class with an actual Android devices, such as Samsung Galaxy S5. Java socket class takes the input in integers from 0-255 as per the ToS Decimal value. If no traffic class is set, the default value CS0 takes place. 


\subsection{Applications in the Wild}
Before we dive into the network performance, it is vital to understand the behavior of already existing mobile  applications. We selected five  audio/video conferencing and a video streaming application, as they would have been the possible candidate applications for using differentiated services intuitively. For VoIP applications, we initiated calls and collected traffic on the caller side with both Android and iOS devices.

We notice that neither the YouTube application nor the server sets the DSCP. On the other hand, two  VoIP applications request for QoS to the network. However, they use two different values. According to the recommended usages table, Viber uses proper value. In contrast, WhatsApp uses a value which is reserved for the network signaling messages, such as DNS requests. All the other applications did not request differentiated services. 
Another aspect of this discussion is that the requested QoS is not bidirectional implicitly. It depends on the sender and the receiver sockets whether they request a particular quality of service from the network or not. The developers have to set the socket properties on both sides (sender and receiver) in order to request similar quality of service. However, requesting the quality of service is one aspect and being given the quality of service is another. This application layer logic when transfers to the network layer, the network policy decides whether the requested quality of service will be allotted  to the application or not.


\subsection{Access Network Behavior}
We first experiment how mobile and WiFi networks treat the packets of the applications with DSCP values. We used Samsung Galaxy S5 with model number as SM-G900F. We developed an Android tool. The tool generates seven consecutive flows to a remove Amazon server in USA. Each flow is marked with a different DSCP value depicted in Figure~\ref{fig:dscpaths}. The flows carry the packets of random size (100-700 Bytes) at 1Hz for 10 seconds.  We investigate how the access networks treat the packets when they first enter into the networks. We select two 4G public mobile networks (MNO\#1 and MNO\#2), an LTE test network (MNO\#T), and a home WiFi gateway.

The initial results for the selected networks are presented in Figure~\ref{fig:dscpaths}. We can see that public MNOs reset the DSCP values to CS0 for any traffic. On the other hand, the LTE test network and home WiFi network let the DSCP marked packets pass as they are, although the  WiFi\#H and MNO\#2 belong to the same network service provider. The figure also demonstrates that the access networks are reseting the DSCP values. 





\subsection{DSCP and Network Performance}

To realize the performance due to marking flows with DSCPs, it is important that a desired DSCP marked flow is compared with a reference or base flow. In this case, the base flow carries the default best effort service request, CS0. The other flow may request AF, EF or other class selectors. We consider two cellular network conditions. Two competitive flows are generated at the same time from the same device. In the other scenario, two flows are generated from the same device, however, separated by a data connectivity reset for 3-5 minutes. 

In this setup, the tool generates two flows with two remote servers according to the above network configurations. We repeated the experiment for six DSCP pairs and the networks presented in Figure~\ref{fig:dscpaths}. Each flow carries 2900 packets of random size (100-700 Bytes) at 10Hz. From the collected traces at the server side, we compare the inter-arrival time of the packets or packet gaps with Kolmogorov-Smirnov (KS) test. With KS test, we validate that the distribution of packet gaps of two flows are different with 95\% confidence level.



\begin{table}[h!]
\begin{center}
\caption{Loss Rates of the flows with different DSCPs.}
{\footnotesize
 \begin{tabular}{| c  | c | c | c | c | c | c |} 
 \hline
 \textbf{Network} & \textbf{AF31} & \textbf{AF33} & \textbf{AF41} & \textbf{AF43} & \textbf{EF} & \textbf{CS6} \\
\hline
 MNO\#1   & 2.11\%  & 4.19\%  & 0\%  & 1.83\%  & 1.83\% & 4.31\%  \\
\hline
 MNO\#2    & 0\%  & 0\%  & 0\%  & 0\%  & 0\% & 0\%  \\
\hline
 WIFI\#H  &  0.15\%  & 0.03\%  & 0.15\%  & 0.19\%  & 0.43\% & 0.03\%  \\
\hline
\end{tabular}}
\label{table:localloss}
\end{center}
\end{table}

The KS test results suggest that the network did not differentiate the flows according to the DSCP values. However, we observed differentiation when two flows are separated by a data disconnection period. In Table~\ref{table:localloss}, we notice that packets with AF31 has lower loss rate than AF33 with MNO\#1. With a lower packet rate of 1Hz we also observed similar behavior, however, the loss rates were lower.

\subsection{Secure Tunnel} Along with the mentioned elements, we also investigated the performance of other network services explicitly used by the end user, such as VPN services.  Our initial intuition was that the traffic flows originating from mobile devices would escape the DSCP value modification by the network operators since the original packets are encrypted. Alternatively, the VPN services would mark the packets with the same DSCP value observed in the IP headers of the original packets. Therefore, the other network entities would see and react according to the original set value. We experimented with both L2 (L2TP IPSec/PSK) and a commercial L3 VPN (OpenVPN) applications on Galaxy S5 for both WiFi (WiFi\#O) and cellular networks (MNO\#1, MNO\#2). With L3 approach, the operating system offers a tap interface for third party applications, such as OpenVPN and others. The device also has it own L2 VPN functionality which requires the remote IP address of a L2 VPN server. From traffic traces, we have found VPN services were reseting to the default DSCP irrespective of the network types. 

\subsection{Summary}
The investigations in this section reveals the diverse behavior of various network elements participating in the end-to-end communication. Our findings are summarized as follows.  
\begin{itemize}

\item Among audio/video conferencing applications, Viber and WhatsApp set the DSCP values.  In the case of video conferencing, Viber uses the highest drop precedence of CS4, AF43. On the other hand, it uses the lowest drop precedence of class CS3, AF31, for voice calls. In the case of switching call mode, Audio to Video, in the middle of a conversation does not affect the initial set value. 

\item WhatsApp uses the maximum value, CS6, for traffic differentiation, which is mostly used by the in-network routing applications or protocols, such as DNS requests. However, IMO, Skype, and Facebook Messenger do not try to take the advantage to \emph{DiffServ} QoS service.

\item Audio/Video streaming applications try to be indifferent as they do not set any value higher than zero. The downlink streaming application, such as YouTube requires setting DSCP at the server sides, as the downlink traffic dominates.

\item Access network and the transport protocol type does not affect the selection of such values by the applications. The commercial access networks nullify the QoS requests irrespective of the request DSCP value and transport protocol. The other network entities, such as L2 and L3 VPNs, also may behave  have similarly.

\item If the access network does not modify the DSCP, the intermediary routers do not change the values. However, similar to ~\cite{nanno}, we also observed that higher packet rates trigger rate control control policies and thus increase packet loss rate which is irrespective of the DSCP used.

\end{itemize}


It is possible that the extensive usage of the default code point or the class selector, CS0, is the result of naive network parameter configuration by the network operators and VPN service providers. Learmonth et al. \cite{Learmonth:2016} investigated   mobile network operators in Europe with protocol messages and found strange DSCP values reconfigured by the operators. 
\section{Net Neutrality} \label{futurenet}

Apart from naive network configuration, the reason for traffic differentiation, even changing the DSCP value,  by the operators could be motivated by the notion of non-neutrality where the content service provider and the network operator together control the QoS.

We have seen that the operators nullify the DSCP requests and push for the best effort service. 
While \emph{DiffServ} enables applications directly requesting the network for a particular QoS, the  networks still may have their own methods to identify the traffic and guide traffic according to their own defined rules without relying on explicit DSCP requests. The networks may apply different policies, such as shaping multimedia traffic and blocking P2P traffic, based on IP address, port number, bandwidth consumed by an application, and the throughput of a flow~\cite{nanno}. Such policies affect the performance of the applications negatively or the QoS. 
Kahkki et al.~\cite{netpolice} found that ISPs differentiate TCP-based traffic. For example, ISPs shape YouTube, Netflix, and other audio/video streaming traffic. However, some operators may choose to provide better service to some specific services, such as HTTP with port number 80. Although the guaranteed QoS is also important for the user generated content with VoIP applications, the non-neutral treatment is very likely and may affect the required QoS as these applications may hinder the profit of the mobile operators. 




\section{QoS in 5G Networks}
The 5G specification~\cite{3gpp.23.501} suggests that mobile devices  receive QoS policies from the operator networks and apply those policies based on the IP or MAC address. The policies including dropping packets, routing packets according to IP/MAC address, and marking  flows with DSCPs. The policies can be governed by the subscription and traffic type. However, the revealed challenges are also applicable to 5G architecture. It is common that operators inherit previous network configurations or simply follow the default configurations from the vendors.
The second challenge is HTTPS and VPN applications. Since the VPN clients encrypt the packets, the network is unable to decrypt those packets. The only solution is that the VPN services abide by the original DSCP requests. In the case of HTTPS traffic, it is unclear how the traffic type will be determined. Although a new draft suggests that the applications add their  signatures at the very beginning of the first packet~\cite{3gpp.23.787}, such approach raises  users' privacy concern and requires API level modification of the mobile systems which makes such method infeasible for practical deployment. Therefore, it is very likely that the future 5G mobile networks will be driven by simple subscription based QoS policies, unless a new deployable solution is in place. Finally, the 5G specification defines QoS  only for the access network and network slicing for the core network. It is yet to be specified how to map QoS policies with the network slices, however, such mapping would depend on operator policies in practice. 


\section{Conclusions}

The seamless QoS depends not only on the applications, but also on the networks and content service providers, and other intermediary middle boxes, such as VPN servers. Selecting appropriate DSCP is also important, as imposing guaranteed QoS may increase cost for the users~\cite{DBLP:journals/corr/abs-1105-0283}. Given the practical limitations described earlier, the most feasible approach would be that the application developers understand the QoS requirements of their applications and set the DSCPs accordingly. Otherwise, the system can take the initiative by the understanding the requirements and set the appropriate DSCP codes.  Similarly, the VPN service providers can adhere to the QoS requirements from the original IP header to maintain the QoS for the tunnels~\cite{rfc2983}. These will further improve the trust on QoS requirements among the participating entities and the trust can propagate along the path.  

However, to accommodate the diverse QoS requirements, like low latency, high data rate and high reliability, DSCP-like  decentralized network layer mechanism may not be sufficient. Since the end-to-end nature of TCP/IP protocol, the QoS requirement of high-level description has to be flexibly mapped to the appropriate infrastructural elements through network layer to physical layer. Therefore,  we may have to devise or design other new approaches in other layers, e.g., logically slicing core and radio network resources \cite{8253541}, and adopting centralized control mechanism, like SDN~\cite{ordonez2017network}, for managing  slices. 


Nevertheless, the barriers are not only technological but also have political implications. The non-neutrality may favor the already existing big players while driving out the neutral or new players from the market. Although 5G aims for competition by allowing innovative business models through customized network services, the challenges remain the same. This is because sharing physical resources among the stakeholders will raise the competition and may affect the QoS. Therefore, there should be some regulations for the content providers and the network service providers to maintain QoS reasonable, and a balanced tariff policy can be applied while new technological advancements take place. 


{\footnotesize \bibliographystyle{acm}
\bibliography{bibliography}}

\vspace{12pt}
\color{red}

\end{document}